\documentclass[%
 reprint,
superscriptaddress,
 showpacs,preprintnumbers,
 amsmath,amssymb,
 aps,
 prc,
floatfix,
fleqn
]{revtex4-1}
\usepackage{graphicx}
\usepackage{mathrsfs}
\usepackage{dcolumn}
\usepackage{bm}
\usepackage[mathlines]{lineno}
\usepackage{xcolor}
\begin{document}

\title{Direct Urca processes involving singlet proton superfluidity in neutron star cooling}

\author{Yan Xu}
\email{Corresponding E-mail: xuy@cho.ac.cn}
\affiliation{Changchun Observatory, National Astronomical Observatories, CAS, Changchun 130117, China}%

\author{Xiu Lin Huang}
\affiliation{Center for Theoretical Physics, Jilin University, Changchun 130012, China}%

\author{Xiao Jun Zhang}
\affiliation{Changchun Observatory, National Astronomical Observatories, CAS, Changchun 130117, China}%

\author{Tmurbagan Bao}
\affiliation{College of Physics and Electronic Information, Inner Mongolia University for the Nationalities, Tongliao 028043, China}%

\author{Lin Xiao}
\affiliation{Changchun Observatory, National Astronomical Observatories, CAS, Changchun 130117, China}%

\author{Cun Bo Fan}
\affiliation{Changchun Observatory, National Astronomical Observatories, CAS, Changchun 130117, China}%

\author{Cheng Zhi Liu}
\email{Corresponding E-mail: lcz@cho.ac.cn}
\affiliation{Changchun Observatory, National Astronomical Observatories, CAS, Changchun 130117, China}%
\date{\today}

\begin{abstract}
A detailed description of the baryon direct Urca processes A: $n\rightarrow p+e+\bar{\nu}_{e}$, B: $\Lambda\rightarrow p+e+\bar{\nu}_{e}$, C: $\Xi^{-}\rightarrow\Lambda+e+\bar{\nu}_{e}$ related to the neutron star cooling is given in the relativistic mean field approximation.
The contributions of the reactions B and C on the neutrino luminosity are calculated by means of the relativistic expressions of the neutrino energy losses.
Our results show that the total neutrino luminosities of the reactions A, B, C within the mass range 1.603-2.067$M_{\odot}$ (1.515-1.840$M_{\odot}$ for TM1 model) for GM1 model are larger than the corresponding values for neutron stars in npe$\mu$ matter.
Although the hyperon direct Urca processes B and C reduce the neutrino emissivity of the reaction A, it illustrates the reactions B and C still make the total neutrino luminosity enhancement in the above mentioned areas.
Furthermore, when we only consider the $^{1}S_{0}$ proton superfluidity in neutron star cooling, we find that although the neutrino emissivity of the reactions A and B is suppressed with the appearance of $^{1}S_{0}$ proton superfluidity, the total contribution of the reactions A, B, C can still quicken a massive neutron star cooling.
These results could be used to help prove appearing hyperons in PSR J1614-2230 and J0348+0432 from neutron star cooling perspective.
\end{abstract}

\maketitle

\section{\label{introduction}Introduction}
Neutron star (NS) constitutes one of the best astrophysical laboratories for studying dense matter physics.
It arises at the end of life of a 8-20 $M_{\odot}$ massive stars and forms in the aftermath of the core collapse supernovae explosion.
A newly born NS is very hot with temperature as high as $10^{11}$-$10^{12}$K, but rapidly cools to a temperature of less than $10^{10}$K within minutes.
The cooling process of a NS is dominated by a combination of surface photon emission and interior neutrino emission.
The latter is responsible for about $10^{5}$-$10^{6}$ years until the interior temperature reaches $10^{6}$K.
It is generally known that photon luminosity is obviously lower than neutrino luminosity, meaning that the thermal radiation from a NS surface reflects the intensity of interior neutrino emission\cite{Yakovlev:APAA42/2004,Yakovlev:AIPC983/2008}.
While neutrino emisision depends strongly on the composition of superdense matter in NSs.
It is well known that NSs cores are dense enough to allow for emerging exotic matter with the strangeness quantum number through weak equilibrium, such as $\Lambda$, $\Sigma^{0}$, $\Sigma^{+}$, $\Sigma^{-}$, $\Xi^{0}$, $\Xi^{-}$ hyperons, referred as nphe$\mu$ matter, except for the conventional nucleons and leptons (npe$\mu$ matter)\cite{Ji:PRD57/1998,Yakovlev:PU42/1999,Zhao:CSB56/2011,Gao:APSS334/2011,Sotani:NPA906/2013,Schaab:APJ504/1998,Wang:PRC81/2010,Xu:RAA15/2015}.
It means that all the possible baryon neutrino emission processes would happen during the neutrino cooling stage \cite{Tsuruta:1964,Flowers:APJ205/1976,Maxwell:APJ231/1979,Gudmundsson:APJ272/1983,Page:APJ394/1992,Kaminker:AA373/2001,Yakovlev:NPA752/2005,Kouvaris:PRD77/2008,Blaschke:PRC85/2012,Beznogov:MNRAS447/2015}.
Among them, the most powerful enhancement of neutrino emission is provided by the nucleon direct Urca processes (NDUP), secondarily is
the hyperon direct Urca processes (HDUP) \cite{Lattimer:PRL66/1991,Prakash:APJ390/1992,Haensel:AA290/1994,Levenfish:AL20/1994,Gusakov:AA389/2002,Xu:CPL28/2011,Xu:CTP56/2011,Xu:CSB59/2014}.
Besides, the degrees of freedom of hyperons tend to soften the equation of state (EOS) calculated in the relativistic mean field (RMF) model based on SU(6) spin-flavor symmetry(quark model for the vector meson-hyperon coupling constants), then reduce the maximum mass of NS to about 1.6-1.7$M_{\odot}$ \cite{Boguta:NPA292/1977,Boguta:2PLB106/1981,Boguta:PLB120/1983,Boguta:PRL59/1987,Schaffner:PRC53/1996,Yang:PRC77/2008,Xu:CPL30/2013}.
However, Demorest et al. in 2010\cite{Demorest:Nature467/2010} indicated that the binary millisecond pulsar PSR J1614-2230 expanded the maximum observational mass from 1.67 $\pm$ 0.02$M_{\odot}$ to 1.97 $\pm$ 0.04$M_{\odot}$ using the Shapiro delay measurements from radio timing observations.
Antoniadis et al. in 2013\cite{Antoniadis:Science340/2013} observed another massive neutron star PSR J0348+0432, whose mass is 2.01 $\pm$ 0.04$M_{\odot}$.
It is clear that the inclusion of hyperons in such heavy NS cores are difficult to explain by SU(6) spin-flavor symmetry in RMF model.
And for this reason, the SU(3) flavor symmetry is widely applied to RMF model. Because it changes the strength of the isoscalar, vector-meson($\omega$ and $\phi$) couplings to the octet states, which can sustain a NS with mass of (1.8-2.1)$M_{\odot}$ even if hyperons exist in NS core \cite{Weissenborn:PRC86/2012,Miyatsu:PRC88/2013,Weissenborn:NPA914/2013,Lopes:PRC89/2014}.
Furthermore, baryons in NS interior can become the superfluid state related to the generation of BB Cooper pairs under attractive interaction.
The baryon superfluidity(SF) could suppress considerably the NDUP, HDUP and thus affect the cooling rate of NS remarkably
\cite{Levenfish:AL20/1994,Yakovlev:PU42/1999,Takatsuka:NPA738/2004}.
As we all know, the neutrons in the crust and protons, hyperons in the core undergo Cooper pair in $^{1}S_{0}$ state, while neutrons in the core can pair in
$^{3}P_{2}$ state.
 \begin{table*}
\centering
\begin{tabular}{c c c c c c c c c c c c c} \hline
&$g_{\sigma N}$&a$(fm^{-1})$&b&$c_{3}$&$g_{\omega N}$&$g_{\rho
N}$&$g_{\phi N}$ &$g_{\sigma\Lambda}$ &$g_{\sigma \Sigma}$
&$g_{\sigma \Xi}$ &$g_{\sigma^{*} \Lambda}$ &$g_{\sigma^{*} \Xi}$\\
\hline
GM1 SU(6) & 9.57& 12.28 &-8.98&0& 10.61 & 4.10& --&5.84&3.87&3.06&3.73&9.67\\
GM1 SU(3) & 9.57& 12.28 &-8.98&0& 10.26 & 4.10& -3.50&7.25&5.28&5.87&2.60&6.82\\
TM1 SU(6) & 10.029& 7.233 &0.618&71.308& 12.614 & 4.632& --&6.17&4.472&3.202&5.015&11.516\\
TM1 SU(3) & 10.029& 7.233 &0.618&81.601& 12.199 & 4.640& -4.164&7.733&6.035&6.328&3.691&8.100\\
\hline
\end{tabular}
\caption[The parameter sets ]{The parameter sets GM1 and TM1. The relations, $g_{\sigma^{*}N}=g_{\rho_{\Lambda}}=0$, are assumed. We take $m_{\omega}=783$MeV, $m_{\rho}=770$MeV, $m_{N}=938$MeV. For the GM1 and TM1 models, $m_{\sigma}=550$MeV and 511.198MeV, respectively.}
\label{tabla1}
\end{table*}
\begin{table*}
\centering
\begin{tabular}{c c c c c c c} \hline
&$g_{\omega \Lambda}$&$g_{\omega \Sigma}$&$g_{\omega \Xi}$&$g_{\phi
\Lambda}$&$g_{\phi \Sigma}$&$g_{\phi \Xi}$
\\ \hline
GM1 SU(6) & 7.073& 7.073 &3.537& 5.002 & 5.002&10.003\\
GM1 SU(3) & 8.149& 8.149 &6.038& -6.253 & -6.253&-9.004\\
TM1 SU(6) & 8.409& 8.409 &4.205& 5.945 & 5.945&11.891\\
TM1 SU(3) & 9.689& 9.689 &7.180& -7.435 & -7.435&-10.706\\
\hline
\end{tabular}
\caption{The other coupling constants for hyperons. The relations, $g_{\rho N}=\frac{1}{2}g_{\rho \Sigma}=g_{\rho \Xi}$, are assumed.}
\label{tabla2}
\end{table*}

This paper is arranged as follows.
In section 2, we make a brief review for RMF, NS cooling theories and the gap equation for the $^{1}S_{0}$ proton SF.
The numerical results are discussed in section 3.
Finally, we summarize our conclusions in section 4.

\section{\label{sec:densityEquations}The density equations}
\subsection{RMF Theory}
In this calculation, we adopt RMF model to describe NS matter.
The constituents of NSs fall into two categories: npe$\mu$ and nphe$\mu$ matter.
The strong interaction between baryons is mediated by the exchange of isoscalar scalar and vector mesons $\sigma$, $\omega$, isovector vector meson $\rho$.
The two additional strange mesons are also included, namely isoscalar scalar $\sigma^{*}$ and vector $\phi$ mesons \cite{Schaffner:PRC53/1996,Yang:PRC77/2008,Xu:CPL29/2012}.
\begin{table*}
\centering
\begin{tabular}{c c c c c} \hline
Processes&Transition&C&$f_{1}$&$g_{1}$\\ \hline
A&$n\rightarrow p+e+\bar{\nu}_{e}$, $p+e\rightarrow n+\nu_{e}$&$\cos\theta_{C}$&1&F+D\\
B&$\Lambda\rightarrow p+e+\bar{\nu}_{e}$, $p+e\rightarrow \Lambda+\nu_{e}$&$\sin\theta_{C}$&$-\sqrt{3/2}$&$-\sqrt{3/2}(F+D/3)$\\
C&$\Xi^{-}\rightarrow\Lambda+e+\bar{\nu}_{e}$, $\Lambda+e\rightarrow\Xi^{-}+\nu_{e}$&$\sin\theta_{C}$&$\sqrt{3/2}$&$\sqrt{3/2}(F-D/3)$\\
D&$\Xi^{-}\rightarrow\Xi^{0}+e+\bar{\nu}_{e}$, $\Xi^{0}+e\rightarrow\Xi^{-}+\nu_{e}$&$\cos\theta_{C}$&1&F-D\\
E&$\Sigma^{-}\rightarrow n+e+\bar{\nu}_{e}$, $n+e\rightarrow \Sigma^{-}+\nu_{e}$&$\sin\theta_{C}$&-1&D-F\\
F&$\Sigma^{-}\rightarrow\Lambda+e+\bar{\nu}_{e}$, $\Lambda+e\rightarrow\Sigma^{-}+\nu_{e}$&$\cos\theta_{C}$&0&$\sqrt{2/3}D$\\
G&$\Sigma^{-}\rightarrow\Sigma^{0}+e+\bar{\nu}_{e}$, $\Sigma^{0}+e\rightarrow\Sigma^{-}+\nu_{e}$&$\cos\theta_{C}$&$\sqrt{2}$&$\sqrt{2}F$\\
H&$\Xi^{-}\rightarrow\Sigma^{0}+e+\bar{\nu}_{e}$, $\Sigma^{0}+e\rightarrow\Xi^{-}+\nu_{e}$&$\sin\theta_{C}$&$\sqrt{1/2}$&$(F+D)/\sqrt{2}$\\
I&$\Xi^{0}\rightarrow\Sigma^{+}+e+\bar{\nu}_{e}$, $\Sigma^{+}+e\rightarrow\Xi^{0}+\nu_{e}$&$\sin\theta_{C}$&1&F+D\\
\hline
\end{tabular}
\caption{The constants of the baryon direct Urca processes. We take $\sin\theta_c$=0.231$\pm$0.003, F=0.477$\pm$0.012, D=0.756$\pm$0.011.} \label{tabla3}
\end{table*}
The total Lagrangian is given by
\begin{eqnarray}
\label{eq:L}
L=\sum_B\overline{\psi}_B[i\gamma_\mu\partial^\mu-(m_B-g_{\sigma B}\sigma-g_{\sigma^*B\sigma^*})\\\nonumber
-g_{\rho B}\gamma_{\mu}{\mathbf{\tau}}\cdot{\mathbf{\rho}^\mu}-g_{\omega B}\gamma_\mu\omega^\mu-g_{\phi B}\gamma_\mu\phi^\mu]\psi_B\\\nonumber
+\frac{1}{2}(\partial_\mu\sigma\partial^\mu\sigma-m_\sigma^2\sigma^2)+\frac{1}{2}(\partial_v\sigma^*\partial^v\sigma^*-m^2_{\sigma^*}\sigma^{*2})\\\nonumber
-\frac{1}{4}W^{\mu v}W_{\mu v}-\frac{1}{4}R^{\mu v}R_{\mu v}-\frac{1}{4}P^{\mu v}P_{\mu v}+\frac{1}{2}m_\omega^2\omega_\mu\omega^\mu\\\nonumber
+\frac{1}{2}m_\rho^2{\mathbf{\rho}}_\mu{\mathbf{\rho}}^\mu+\frac{1}{2}m^2_{\phi}\phi_\mu\phi^\mu+\frac{1}{4}c_3(\omega_\mu\omega^\mu)^2\\\nonumber
-\frac{1}{3}a\sigma^{3}-\frac{1}{4}b\sigma^4+\sum_l\overline{\psi}_l[i\gamma_\mu\partial^\mu-m_l]\psi_l.
\end{eqnarray}
Here $W_{\mu v}=\partial_\mu\omega_v-\partial_v\omega_\mu$,
$R_{\mu v}=\partial_\mu{\mathbf{\rho}}_v-\partial_v{\mathbf{\rho}}_\mu$ and $P_{\mu v}=\partial_\mu\phi_v-\partial_v\phi_\mu$ denote the field tensors of $\omega$, $\rho$ and $\phi$ mesons, respectively.
The sum on B and l runs over the octet baryons and leptons, namely, n, p, $\Lambda$, $\Sigma^{0}$, $\Sigma^{+}$, $\Sigma^{-}$, $\Xi^{0}$, $\Xi^{-}$, e, $\mu$.
$\psi_{B}$, $\psi_{l}$ and $m_{B}$, $m_{l}$ are the baryon, lepton Dirac fields and masses, respectively.
$\gamma_{u}$ is the Dirac matrice.
The meson fields are replaced by their expectation values at the mean field level.
Now we are going to plug the above Lagrangian into the Euler-Lagrange equations
\begin{eqnarray}
\label{eq: EL}
\frac{\partial L}{\partial \psi(x)}-\partial_{\mu}\frac{\partial L}{\partial(\partial_{\mu}\psi)}=0
\end{eqnarray}
The equations of motion for each baryon and meson fields can be obtained in RMF approximation
\begin{eqnarray}
\label{eq: baryon field}
(i\gamma_{\mu}\partial^{\mu}-m_{B}^{*}-g_{\omega
B}\gamma_{0}\omega^{0}-g_{\rho B}\gamma_{0}\tau_{3}\rho^{0}_{3}
\\
-g_{\phi B}\gamma_{0}\phi^0)\psi_{B}=0,
\nonumber
\end{eqnarray}
\begin{eqnarray}
\label{eq: meson field}
\sum_B g_{\sigma B}\rho_{SB}=m_\sigma^2\sigma+a\sigma^2+b\sigma^3,\\
\sum_B g_{\omega B}\rho_B=m_\omega^2\omega_0+c_{3}\omega^{3}_{0},\\
\sum_B g_{\rho B}\rho_{B}I_{3B} =m_{\rho}^2\rho_{03},\\
\sum_B g_{\sigma^* B}\rho_{SB}=m_{\sigma^*}^2\sigma^*,\\
\sum_B g_{\phi B}\rho_{B}=m_\phi^2\phi_0.
\end{eqnarray}
Here $J_{B}$ and $I_{3B}$ express the baryon spin and isospin projections, respectively. $m_B^{*}$ is the baryon effective mass
\begin{eqnarray}
\label{eq: baryon field}
m_B^{*}=m_B-g_{\sigma B}\sigma-g_{\sigma^*B\sigma^*},
\end{eqnarray}
The scalar density $\rho_{SB}$ and baryon density $\rho_{B}$ are given by
\begin{eqnarray}
\label{eq: density}
\rho_{SB}=\frac{2J_{B}+1}{2\pi^{2}}\int_0^{p_{FB}}\frac{m_{B}^{*}}{\sqrt{p_{B}^{2}+m_{B}^{*2}}}p_{B}^{2}dp_{B}\\
\rho_B=\frac{1}{\pi^2}\int_0^\infty dp_{B}p_{B}^2.
\end{eqnarray}
The hadron phase should meet the local charge neutrality and beta-equilibrium conditions.
The former is given by
\begin{eqnarray}
\label{eq: neutrality}
\rho_{p}+\rho_{\Sigma^{+}}=\rho_{\Sigma^{-}}+\rho_{\Xi^{-}}+\rho_{e}+\rho_{\mu}.
\end{eqnarray}
The latter is imposed by the baryon chemical potential, which is a linear combination of $\mu_{n}$ and $\mu_{e}$,
\begin{eqnarray}
\label{eq: chemical potential}
\mu_{B}=\mu_{n}-q_{B}\mu_{e},\mu_{e}=\mu_{\mu}
\end{eqnarray}
where $q_{B}$ is the baryon electric charge(in unit of e).

We can solve the Eqs. (3)-(13) self-consistently at a given baryon density $\rho_{B}$.

The total energy density and pressure of NS matter are
\begin{eqnarray}
\label{eq: energy}
\varepsilon=\sum_B\frac{1}{\pi^{2}}\int_0^{p_{FB}}\sqrt{p_{B}^{2}+m_{B}^{*2}}p_{B}^{2}dp_{B}+\frac{1}{2}m_{\sigma}^{2}\sigma^{2}+\frac{1}{3}a\sigma^{3}\\\nonumber
+\frac{1}{4}b\sigma^4+\frac{1}{2}m_{\sigma^{*}}^{2}\sigma^{*2}+\frac{1}{2}m_{\omega}^{2}\omega^{2}+\frac{3}{4}c_{3}\omega^{4}+\frac{1}{2}m_{\phi}^{2}\phi^{2}\\\nonumber
+\frac{1}{2}m_{\rho}^{2}\rho^{2}+\sum_l\frac{1}{\pi^{2}}\int_0^{p_{Fl}}\sqrt{p_{l}^{2}+m_{l}^{*2}}p_{l}^{2}dp_{l}
\end{eqnarray}

\begin{eqnarray}
\label{eq: press}
P=\frac{1}{3}\sum_B\frac{1}{\pi^{2}}\int_0^{p_{FB}}\frac{p_{B}^{4}dp_{B}}{\sqrt{p_{B}^{2}+m_{B}^{*2}}}-\frac{1}{2}m_{\sigma}^{2}\sigma^{2}-\frac{1}{3}a\sigma^{3}\\\nonumber
-\frac{1}{4}b\sigma^4-\frac{1}{2}m_{\sigma^{*}}^{2}\sigma^{*2}+\frac{1}{2}m_{\omega}^{2}\omega^{2}+\frac{1}{4}c_{3}\omega^{4}+\frac{1}{2}m_{\phi}^{2}\phi^{2}\\\nonumber
+\frac{1}{2}m_{\rho}^{2}\rho^{2}+\frac{1}{3}\sum_l\frac{1}{\pi^{2}}\int_0^{p_{Fl}}\frac{p_{l}^{4}dk}{\sqrt{p_{l}^{2}+m_{l}^{*2}}p_{l}^{2}}dp_{l}
\end{eqnarray}
Eqs. (14) and (15) as inputs, we can obtain the mass-radius relation by solving the Tolman-Oppenheimer-Volkoff(TOV) equation \cite{Oppenheimer:PR55/1939,Tolman:PR55/1939}
\begin{eqnarray}
\label{eq: TOV} \frac{dP(r)}{dr} =
-\frac{[P(r)+\varepsilon(r)][M(r)+4\pi r^{3}P(r)]}{r(r-2M(r))},
\\\nonumber
\frac{dM(r)}{dr}=4\pi r^{2}\varepsilon(r).\\\nonumber
\end{eqnarray}
We adopt two successful RMF parameter sets to describe NS matter, GM1 and TM1, as listed in Table I \cite{Miyatsu:PRC88/2013}.
These parameters have been determined by fitting to some ground state properties of nuclear matter.
As for the couplings of the isoscalar vector mesons $\omega$ and $\phi$ to baryons, we adopt two relations: SU(6) spin-flavor
symmetry based on the naive quark model and general SU(3) flavor symmetry, as listed in Table II \cite{Weissenborn:NPA914/2013}.

\subsection{NS cooling theory}
\label{sec:cooling}
The baryon direct Urca processes consist of two successive reactions, beta decay and capture, are listed in Table III \citep{Prakash:APJ390/1992}.
\begin{eqnarray}
B_{1}\rightarrow B_{2} + e +\bar{\nu}_{e}, B_{2}  + e\rightarrow
B_{1}+ \nu_{e}.
\end{eqnarray}
Here $B_{1}$ and $B_{2}$ represent baryons.
Due to the EOSs of NSs matter are derived by RMF model, so the neutrino energy losses must be
consistent with the used relativistic EOSs.
In the free relativistic gas, the energy and momentum conservations require a large effective mass differece of $B_{1}$ and $B_{2}$, $m_{B_{1}}^{*} - m_{B_{2}}^{*}\sim100$MeV, which is unlikely to appear in the reactions A, D and G. The reason is that the effective masses of hyperons with the same species but the different isospins are same in Eq.(9).
Therefore, in the relativistic regime, the energy conservation should be assured by considering the potential energy difference of $B_{1}$ and $B_{2}$.
The neutrino emissivity can be given by the Fermi Golden Rule
\begin{eqnarray}
Q_{0}=2\int[\prod\limits_{j=1}^{4}\frac{d^{3}p_{j}}{(2\pi)^{12}(2\varepsilon_{j})}]\varepsilon_{4}f_{1}(1-f_{2})(1-f_{3})|M_{fi}|^{2}\\
\times(2\pi)^{4}\delta(E_{1}-E_{2}-\varepsilon_{3}-\varepsilon_{4})\delta\mathbf{(p_{1}-p_{2}-p_{3}-p_{4})}\nonumber,
\end{eqnarray}
where ${p}_{j}$, $\varepsilon_{j}$ express the momentum and kinetic energy of particle species j ($j = 1, 2, 3$ and 4 refer to $B_{1}$,
$B_{2}$, e and $\bar{\nu}_{e}$), respectively.
$f_{j}$ is the Fermi-Dirac distribution functions of baryons and electrons,
\begin{eqnarray}
f_{B}=\frac{1}{\exp((E_{B}-\mu_{B})/T)+1},\\\nonumber
f_{e}=\frac{1}{\exp((\varepsilon_{3}-\mu_{e})/T)+1}.
\end{eqnarray}
The delta functions $\delta(E_{1}-E_{2}-\varepsilon_{3}-\varepsilon_{4})$ and $\delta\mathbf{(p_{1}-p_{2}-p_{3}-p_{4})}$ describe the energy and momentum conservation.
$E_{1, 2}=\varepsilon_{1, 2}+U_{1, 2}$ is the energy of baryons.
$U_{1, 2}$ is the potential energy of baryons, which can be obtained in Section A and has the following form
\begin{eqnarray}
U_{B}=g_{\omega B}\omega_{0}+g_{\rho B}I_{3B}\rho_{0}+g_{\phi B}\phi_{0}.
\end{eqnarray}
Namely,
\begin{eqnarray}
U_{n}-U_{p}=-g_{\rho N}\rho_{0},
\\\nonumber
U_{\Sigma^{-}}-U_{\Sigma^{0}}=-g_{\rho \Sigma}\rho_{0},
\\\nonumber
U_{\Xi^{-}}-U_{\Xi^{0}}=-g_{\rho \Xi}\rho_{0}.
\end{eqnarray}

$|M_{fi}|^{2}$ is the squared matrix element of the baryon direct Urca processes summed over spins of initial and final particles
\begin{eqnarray}
|M_{fi}|^{2}=32G_{F}^{2}C^{2}[(g_{1}^{2}-f_{1}^{2})M^{*}_{1}M^{*}_{2}(P_{4}P_{3})+(g_{1}\\
-f_{1})^{2}(P_{4}P_{2})(P_{3}P_{1})+(g_{1}+f_{1})^{2}(P_{4}P_{1})(P_{3}P_{2})]\nonumber,
\end{eqnarray}
where $P_{j}=(\varepsilon_{j},\mathbf p_{j})$. $G_{F}=1.436\times10^{-49}$ erg cm$^{3}$ is the weak-coupling constant.
$f_{1}$, $g_{1}$ and C are the vector, axial-vector constants and Cabibbo angle which are given in Table III.

The relativistic expression of the energy loss Q per unit volume and time in NS matter is expressed as \cite{Leinson:PLB518/2001,Leinson:NPA707/2002}
\begin{eqnarray}
&Q&=\frac{457\pi}{10080}G_{F}^{2}C^{2}T^{6}\Theta(p_{F3}+p_{F2}-p_{F1})\\
&\times&\{f_{1}g_{1}((\varepsilon_{F_{1}}+\varepsilon_{F_{2}})p_{F3}^{2}-(\varepsilon_{F_{1}}-\varepsilon_{F_{2}})(p_{F1}^{2}-p_{F2}^{2}))\nonumber
\\
&+&2g_{1}^{2}\mu_{e}m_{1}^{*}m_{2}^{*}+(f_{1}^{2}+g_{1}^{2})(\mu_{e}(2\varepsilon_{F_{1}}\varepsilon_{F_{2}}-m_{1}^{*}m_{2}^{*})\nonumber
\\
&+&\varepsilon_{F_{1}}p_{3}^{2}-\frac{1}{2}(p_{F1}^{2}-p_{F2}^{2}+p_{F3}^{2})(\varepsilon_{F_{1}}+\varepsilon_{F_{2}}))\}\nonumber,
\end{eqnarray}
In this expression,
$p_{F1}$, $p_{F2}$ and $p_{F3}$ are the Fermi momenta of baryons and leptons.
$\varepsilon_{F_{1}}$ and $\varepsilon_{F_{2}}$ are the kinetic energy of baryon at the Fermi surface.
$\Theta = 1$ if the Fermi momenta $p_{F1}$, $p_{F2}$, $p_{F3}$ satisfy the triangle condition and $\Theta = 0$ otherwise.
The situation of muons is similar to that of electrons.

The cooling equation based on the approximation of isothermal interior is,
\begin{eqnarray}
\label{eq: Cool} C_{v}\frac{dT}{dt}=-(L_\nu+L_r).
\end{eqnarray}
Here $L_{\nu}$ and $L_{r}$ are the total neutrino and photon luminosities, respectively.
$C_{v}$ is the total thermal capacity of NS matter.
They are
\begin{eqnarray}
\label{eq: Luminosity}
L_{\nu}=\int Q_{0} e^{2\Phi}dV, L_{r}=4\pi R^{2}\sigma (10T)^{\frac{8}{3}}e^{2\Phi_{s}},
\\\nonumber
C_{v0}=\int(C_{e}+C_{\mu}+\sum_B C_{B0})dV,
\end{eqnarray}
where $\sigma$ is the Stefan-Boltzmann constant, $e^\Phi=\sqrt{1-2m/r}$ is the gravitational redshift.
$e^{\Phi_{s}}$ is the value of $e^\Phi$ at the stellar surface (r=R).
The semiempirical expression $T_{s}=(10T)^{2/3}$ expresses the relation between interior temperature T and surface temperature $T_{s}$.

\subsection{SF of protons}
\label{sec:SF}
The key quantity in determining the onset of $^{1}S_{0}$ proton SF is the gap function $\Delta(p)$,
\begin{equation}
\Delta(p)=-\frac{1}{4\pi^{2}}\int{p^{'2}dp^{'}\frac{V(p,p^{'})\Delta(p^{'})}{\sqrt{\varepsilon^{2}(p^{'})+\Delta^{2}(p^{'})}}},
\end{equation}
where $\varepsilon(p)=E(p)-E(p_{Fp})$.
$E(p)$ is the single-particle energy of protons with momentum p
\begin{equation}
E(p)=\sqrt{p^{2}+m_{p}^{*2}}+g_{\omega p}\omega+g_{\phi p}\phi+g_{\rho p}I_{3p}\rho.
\end{equation}
$V(p,p^{'})$ is the pp potential matrix element.
In this work, we  use the Reid soft core(RSC) potential for the pp potential\cite{Sprung:NPA168/1971,Amundsen:NPA437/1985,Nishizaki:PTP86/1991,Wambach:NPA555/1993}, as an example to demonstrate the influence of hyperons on the $^{1}S_{0}$ proton pairing energy gaps.
%
%
%
\begin{figure}
\includegraphics[width=3.2in,height=4.8in]{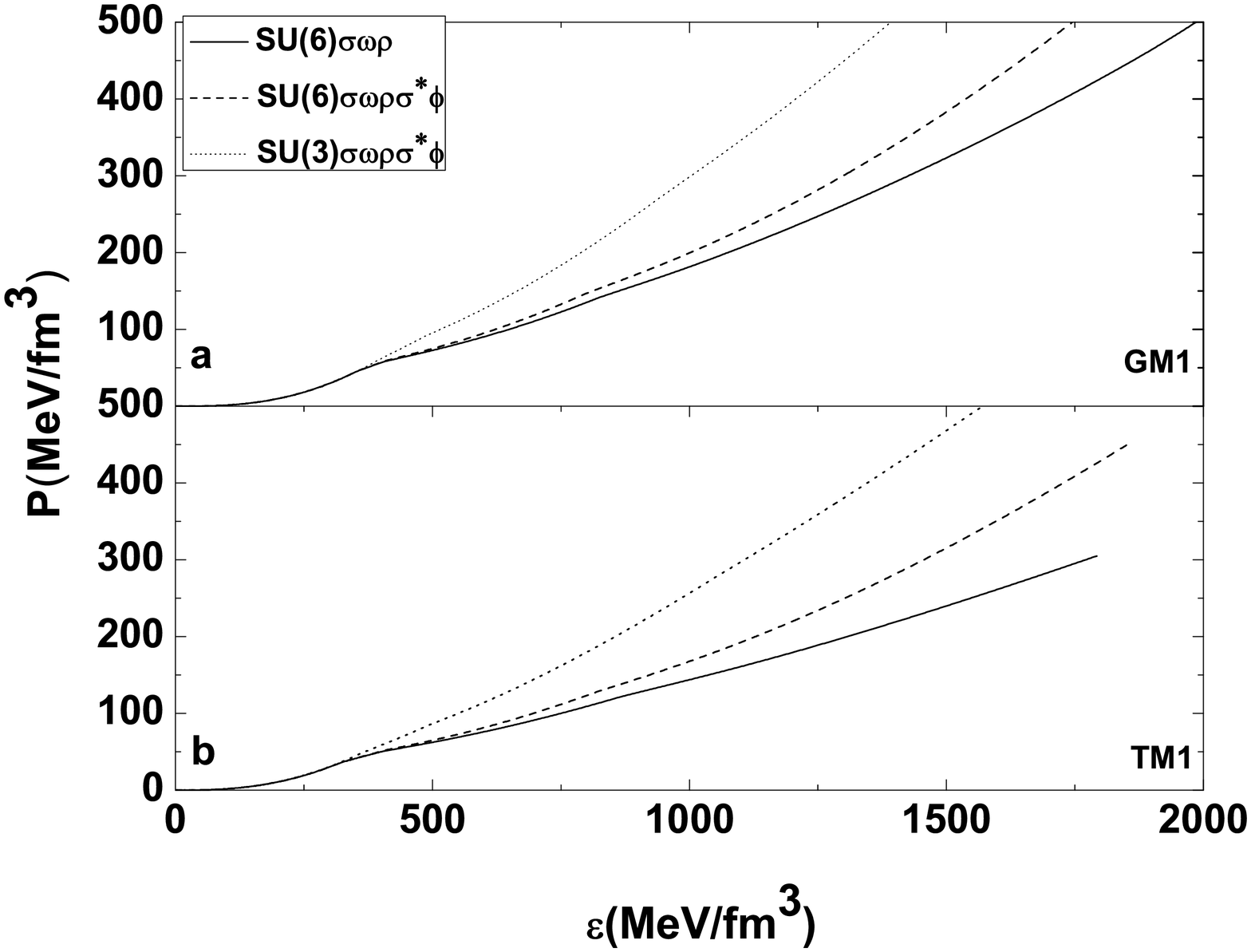}
\caption[]{EOSs including hyperons in NS matter.}
\label{fig:fig1}
\end{figure}
\begin{figure}
\includegraphics[width=3.2in,height=4.8in]{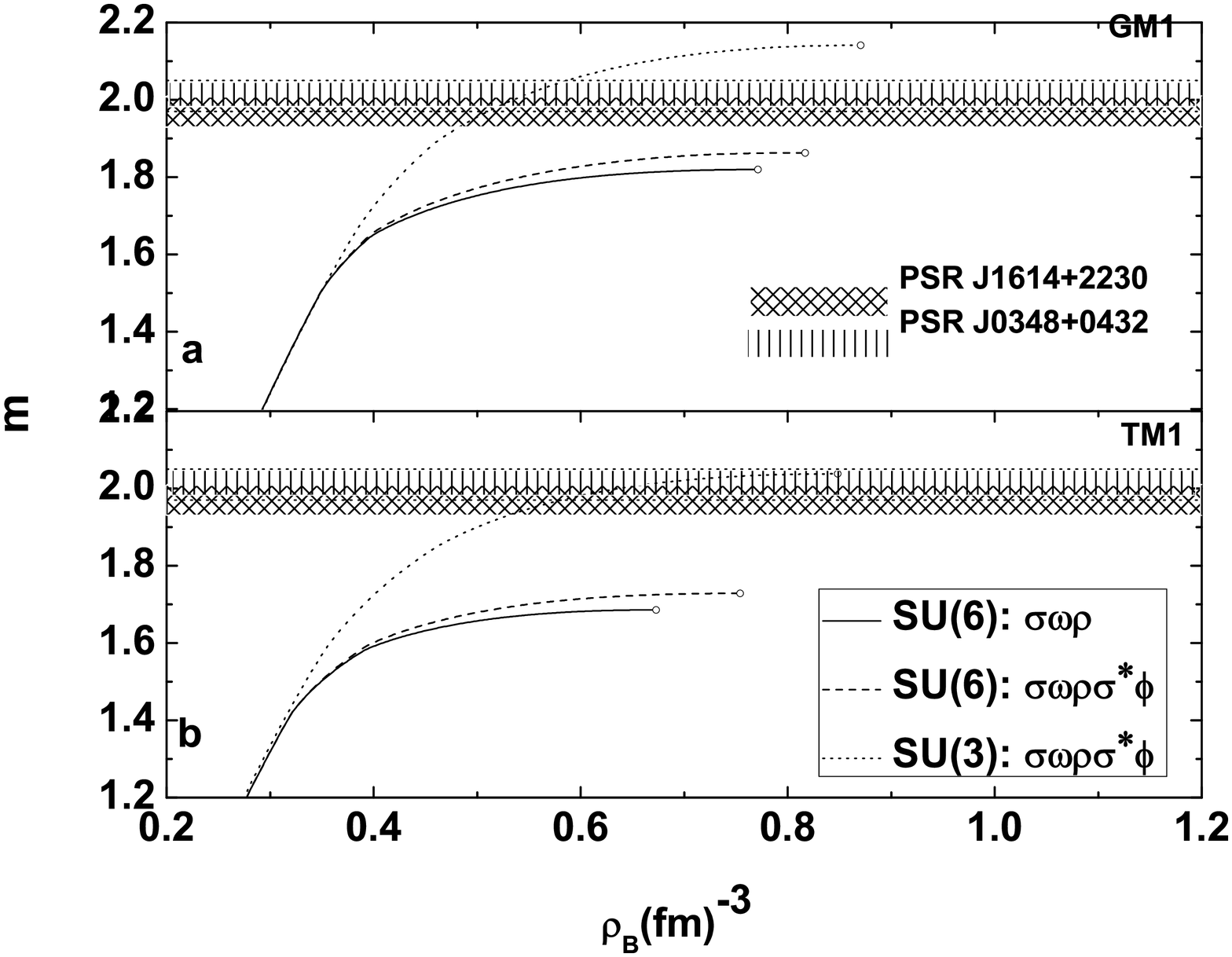}
\caption[]{Mass-radius relations including hyperons.}
\label{fig:fig2}
\end{figure}
The critical temperature $T_{cp}$ of $^{1}S_{0}$ proton SF is given by the pairing gap $\Delta(k)$ at zero temperature approximation,
\begin{equation}
T_{cp}\doteq 0.57 \Delta(k).
\end{equation}
\begin{figure}
\includegraphics[width=3.8in,height=4.8in]{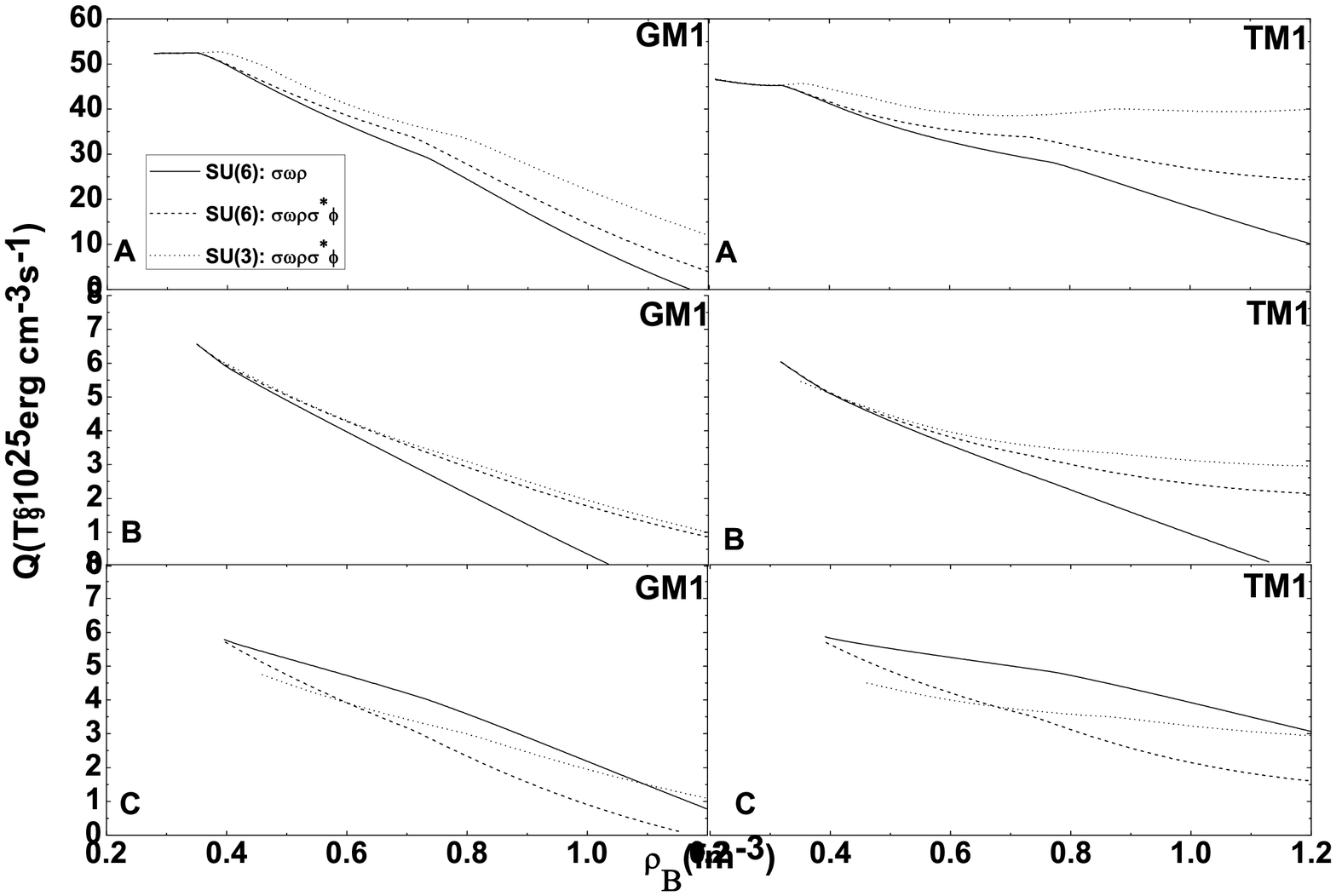}
\caption[]{Neutrino emissivities of the reactions A, B and C as a function of the baryon density $\rho_{B}$ in nphe$\mu$ matter.}
\label{fig:fig3}
\end{figure}
\begin{figure}
\includegraphics[width=3.8in,height=4.2in]{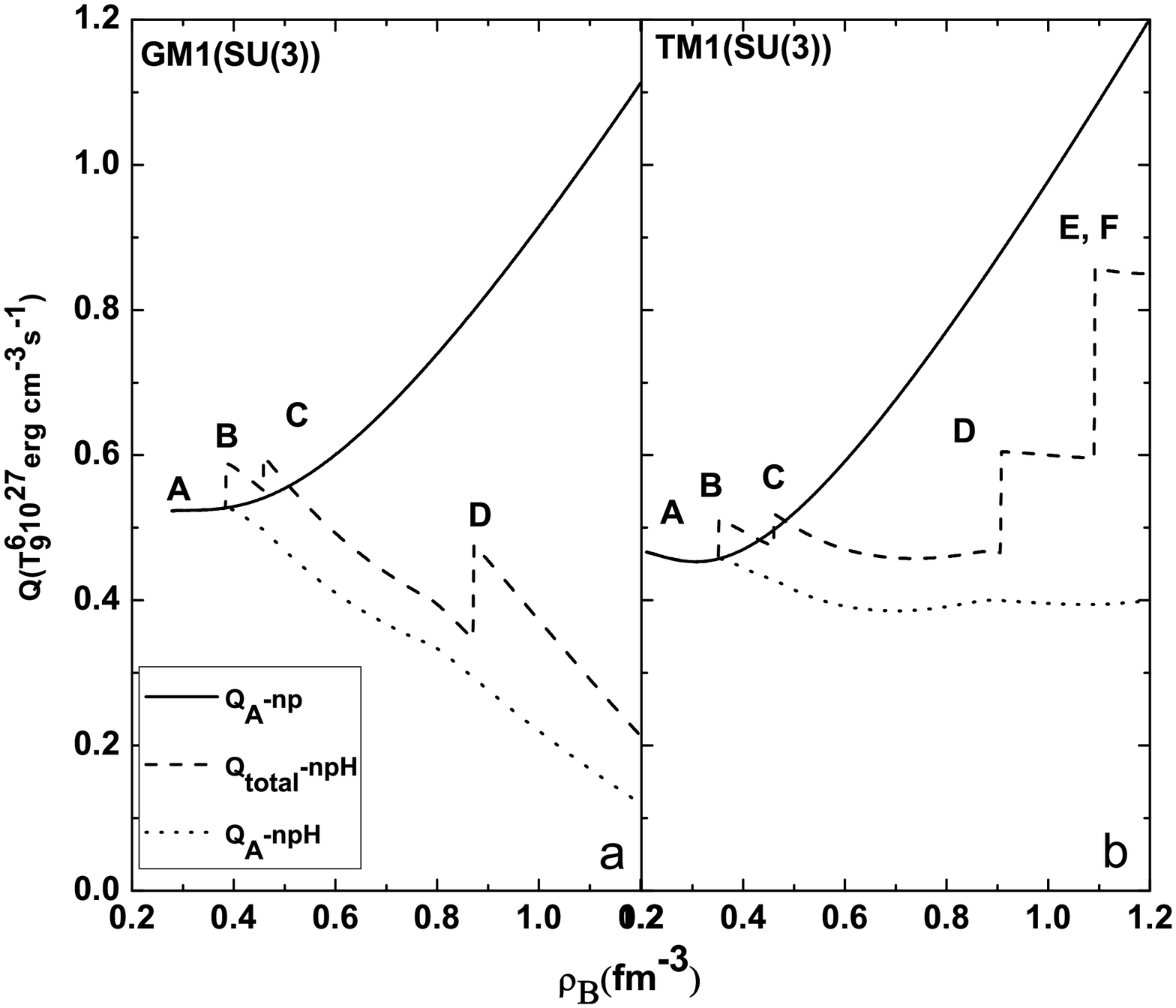}
\caption[]{Total neutrino emissivities of the reactions A-F as a function of the baryon density $\rho_{B}$. The solid line is the neutrino emissivity of the reaction A in npe$\mu$ matter. The dashed line is the total neutrino emissivity of the reactions A-F in nphe$\mu$ matter. The dotted line is the neutrino emissivity of the reaction A in nphe$\mu$ matter.}
\label{fig:fig4}
\end{figure}
\begin{figure}
\includegraphics[width=3.8in,height=4.8in]{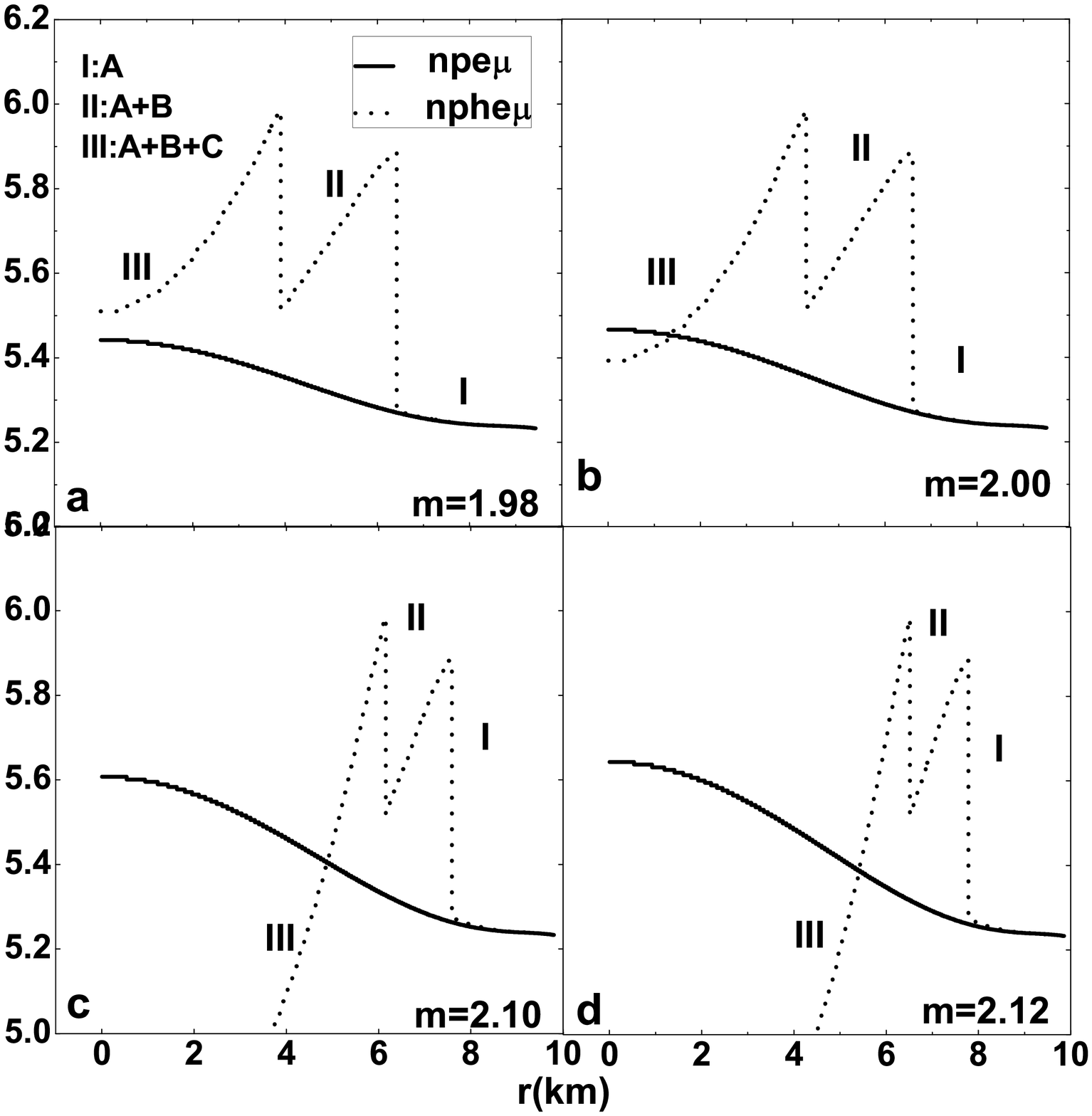}
\caption[]{Radial distributions of the total neutrino emissivities with different
mass NSs in npe$\mu$ matter(solid lines) and nphe$\mu$ matter(dotted lines) in the GM1 model}\label{fig:fig5}
\end{figure}
\begin{figure}
\includegraphics[width=3.8in,height=4.2in]{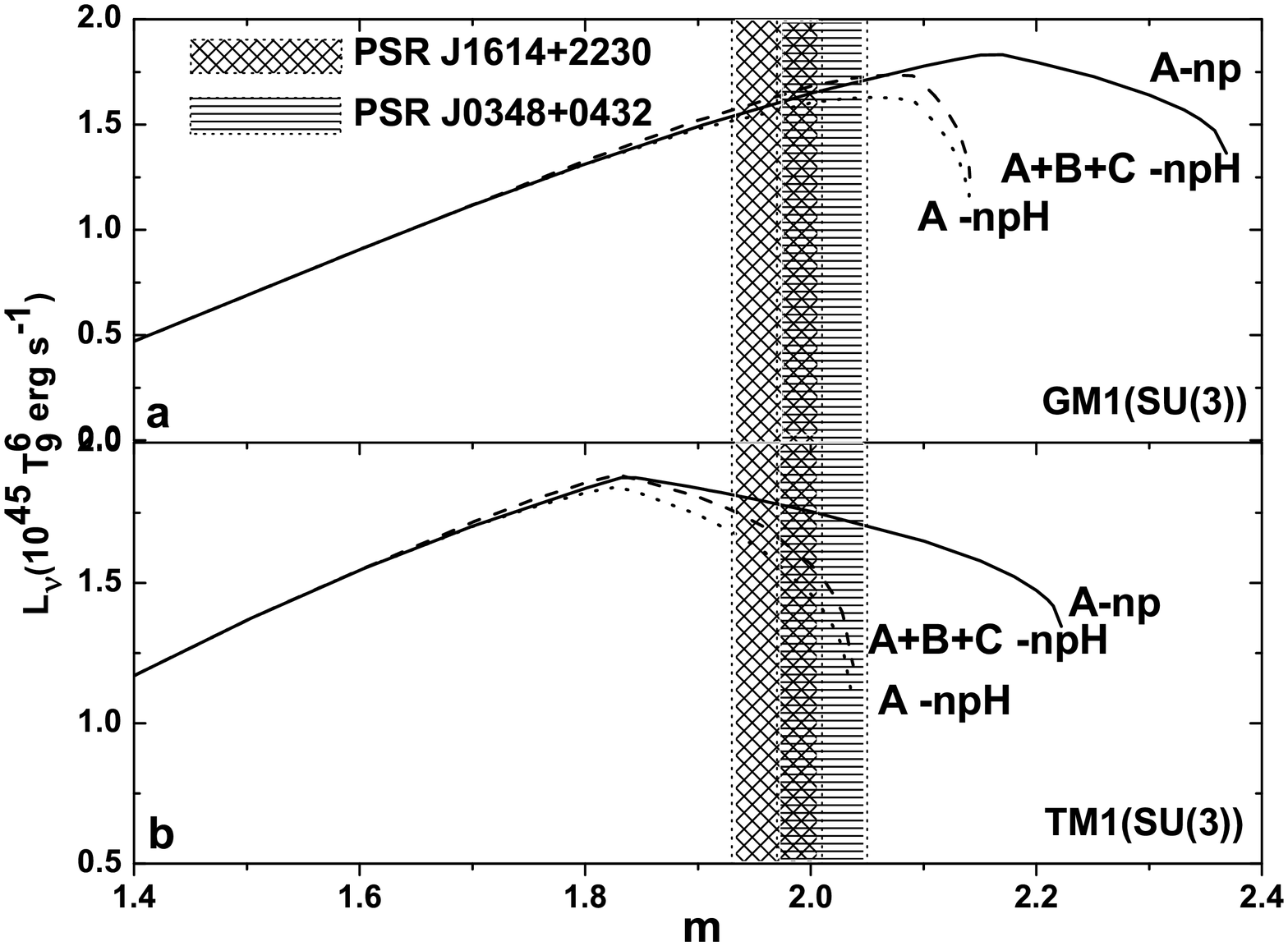}
\caption[]{Total neutrino luminosity of the reactions A, B, C as a function of the NS mass m. The solid line is the neutrino luminosity of the reaction A in npe$\mu$ matter. The dashed line is the total neutrino luminosity of the reactions A, B, C in nphe$\mu$ matter. The dotted line is the neutrino luminosity of the reaction A in nphe$\mu$ matter.} \label{fig:fig6}
\end{figure}
\begin{figure}
\includegraphics[width=3.8in,height=4.8in]{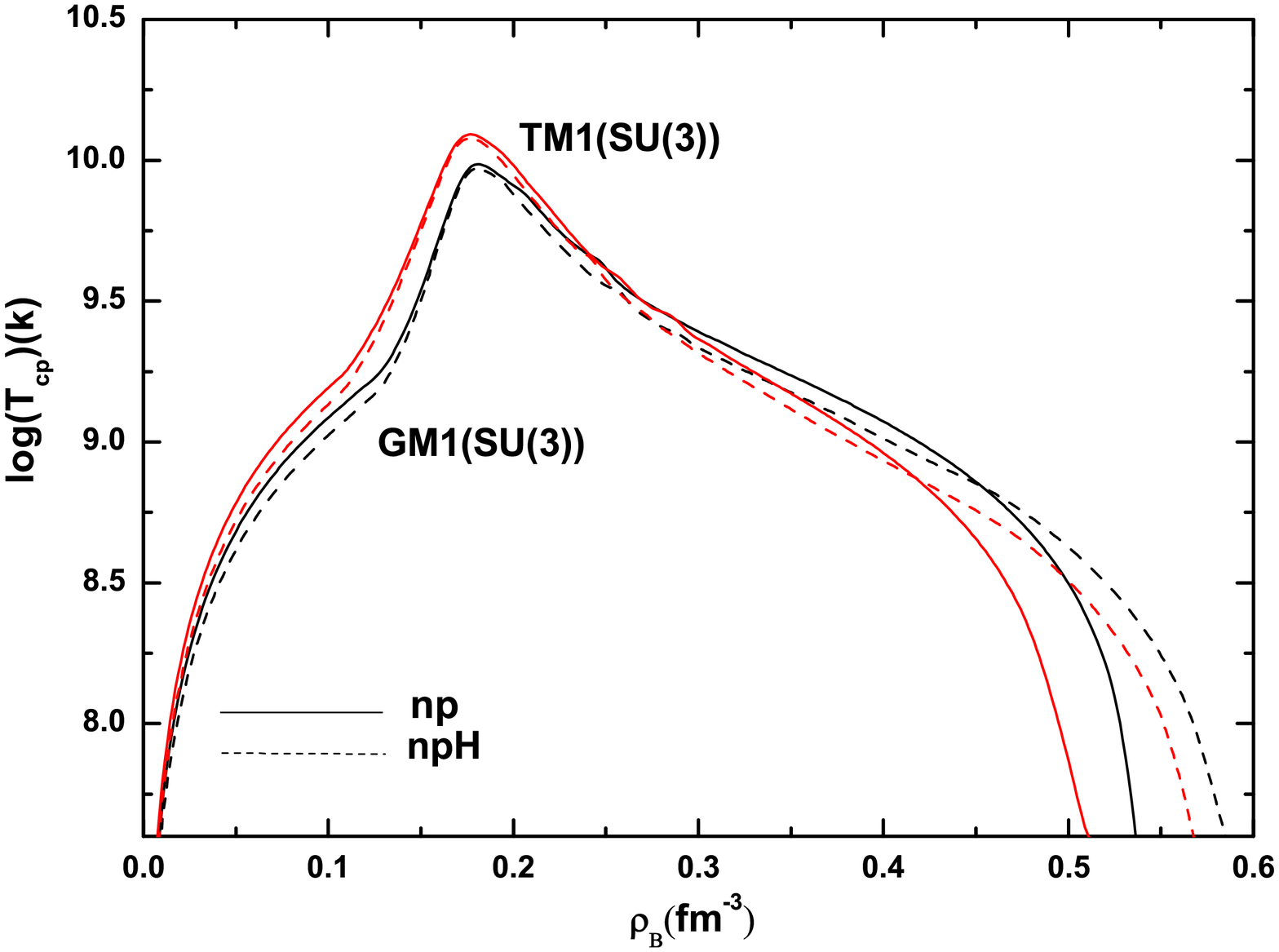}
\caption[]{The critical temperature $T_{cp}$ of the $^{1}S_{0}$ proton SF as a function of the NS mass m in npe$\mu$ matter(solid lines) and nphe$\mu$ matter(dashed
lines)}\label{fig:fig7}
\end{figure}
As a result, the neutrino emissivity and thermal capacity can be written as
\begin{equation}
Q=Q_{0}R_{B},
\\
C_{B}=C_{B0}R_{C_{B0}}.
\end{equation}
Here $R_{B}$ and $R_{C_{B0}}$ are the SF reduction factors of the neutrino emissivity and thermal capacity, respectively.
For the $^{1}S_{0}$ proton SF, the reduction factors $R_{p}$ and $R_{C_{p0}}$ are
\begin{equation}
R_{p}=\frac{0.0163exp(\frac{-1.764T_{cp}}{T})}{(\frac{T}{T_{cp}})^{5.5}},
\\
R_{C_{p0}}=\frac{3.149exp(\frac{-1.764T_{cp}}{T})}{(\frac{T}{T_{cp}})^{2.5}}.
\end{equation}
According to the discussion of the RMF approach above, we can obtain the EOS, mass-radius relations and neutrino emissivities of the reactions A, B, C as well as Fermi momentum and single particle energy of protons, then the pairing gap and critical temperature of $^{1}S_{0}$ proton SF and speed of the NS cooling can be obtained.
\begin{figure}
\includegraphics[width=3.8in,height=4.2in]{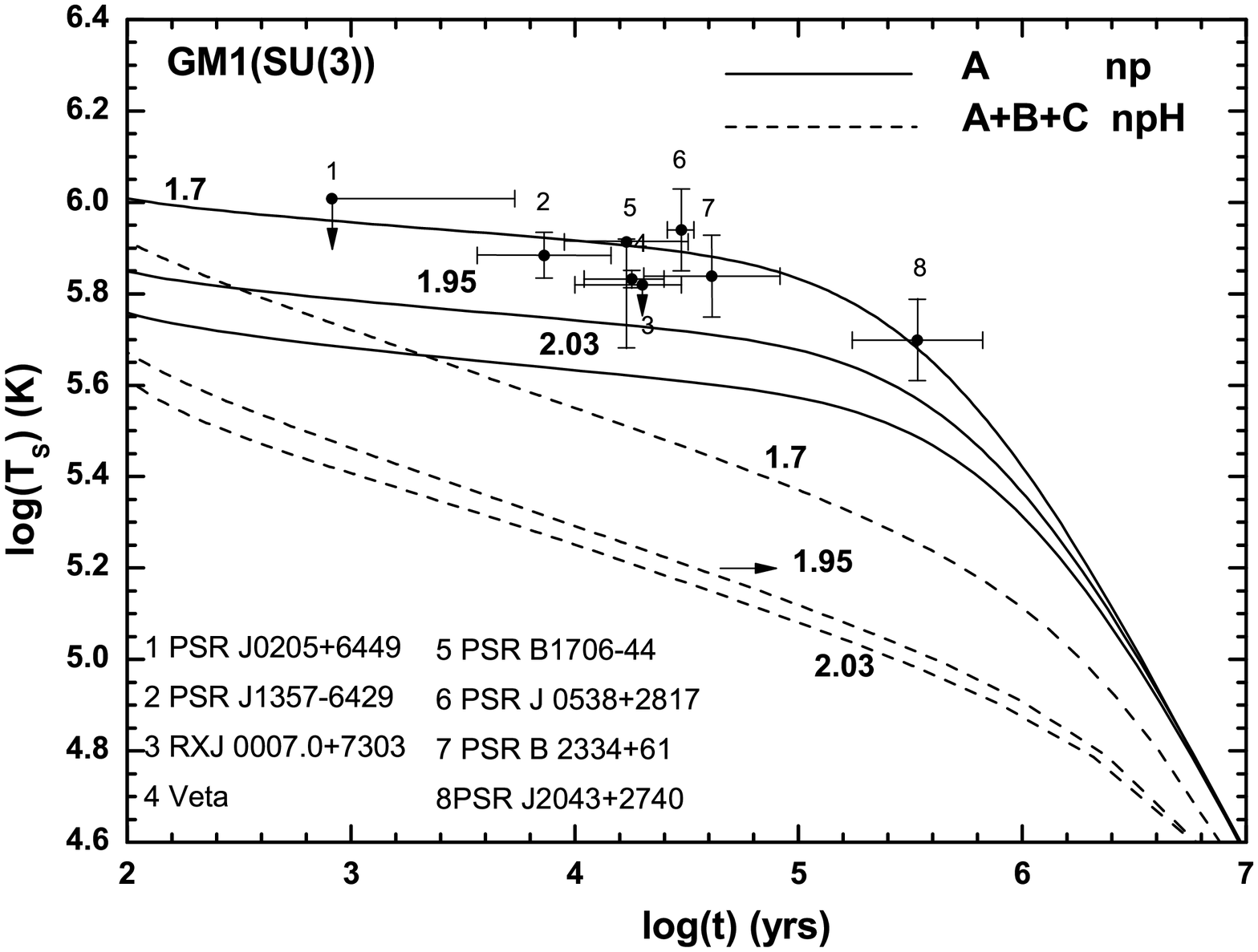}
\caption[]{Observational data(error bars) on surface temperatures of 8 NSs as compared with theoretical cooling curves obtained by the GM1 model for proton SF from Fig.7. The solid lines correspond to npe$\mu$ matter, the dashed lines correspond to nphe$\mu$ matter with masses(from top to bottom)1.5, 1.7, 1.95 and 2.03$M_{\odot}$, respectively.}
\label{fig:fig8}
\end{figure}
\begin{figure}
\includegraphics[width=3.8in,height=4.8in]{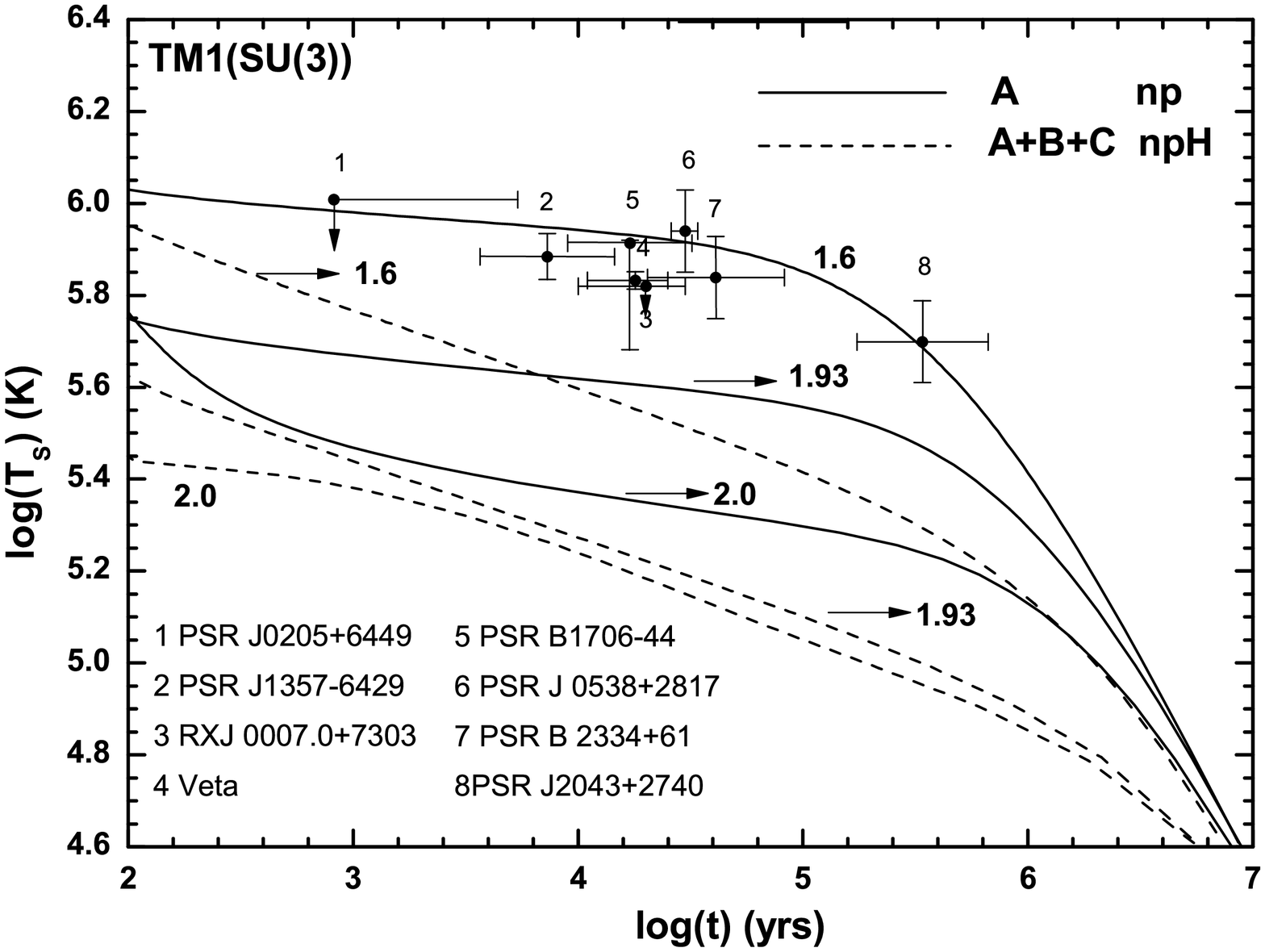}
\caption[]{Observational data(error bars) on surface temperatures of 8 NSs as compared with theoretical cooling curves obtained by the TM1 model for proton SF from Fig.7. The solid lines correspond to npe$\mu$ matter, the dashed lines correspond to nphe$\mu$ matter with masses(from top to bottom) 1.4, 1.6, 1.93 and 2.0$M_{\odot}$, respectively.}
\label{fig:fig9}
\end{figure}
\section{\label{sec:Results} Results and Discussion}
In this section, we give three cases in Eq.(1) for RMF theory: (i) the non-strange $\sigma$, $\omega$, $\rho$ mesons are included
in SU(6) spin-flavor symmetry; (ii) the $\sigma$, $\omega$, $\rho$ mesons including strange mesons $\sigma^{*}$ and $\phi$ are considered in SU(6) spin-flavor symmetry; (iii) $\sigma$, $\omega$, $\rho$, $\sigma^{*}$ and $\phi$ mesons are taken into account in SU(3) flavor symmetry.
We mainly study the effects of the degrees of freedom of hyperons and reactions B, C on the EOS,
neutrino emissivity, neutrino luminosity, energy gap of $^{1}S_{0}$ proton SF and NS cooling.
Then we compare our results with PSR J1614-2230 and J0348+0432, whose measured masses are used as reference values.

Fig. 1 shows the EOSs in three cases.
Fig. 2 shows the mass-radius relations of NSs by solving the TOV equation.
The softest and hardest EOSs are obtained by cases (i) and (iii), respectively.
Though the coupling $g_{\omega N}$ for case (iii) is smaller than the corresponding value for case (i) as shown in Table 1, the total repulsive force is attributed not only to $\omega$ meson but also to $\phi$ meson.
As seen in Fig. 1 and 2, though we consider the contribution of the strange mesons $\sigma^{*}$ and $\phi$ on the EOS in case (ii), the coupling $g_{\phi N}$=0.
It means that $\phi$ meson only couples to hyperons and makes the EOS be not enough stiff.
So the hardest EOS is obtained only through the $\phi$ meson in case (iii).
From case (i) to (iii), the maximum mass of NS (the corresponding center density) sequently increases from 1.820 (0.771), 1.863 (0.817) to 2.141$M_{\odot}$ (0.871) for the GM1 model, 1.686 (0.673),
1.729 (0.754) to 2.038$M_{\odot}$ (0.848) for the TM1 model, respectively (Fig. 2).
Namely, the EOS in SU(3) flavor symmetry could be consistent with the observed values of PSR J1614-2230 and J0348+0432 when hyperons appear in NS core.
Fig. 3 depicts the neutrino emissivities of the reactions A, B, C in nphe$\mu$ matter for the three cases.
In order to make the effects of hyperons and reactions B, C on A more intuitive, Fig. 4 depicts the total neutrino emissivity of the reactions A, B, C in npe$\mu$ and nphe$\mu$ matter for case (iii).
As can be seen from Figs. 3 and 4, the neutrino emissivity of the reaction A has a tendency to decrease with increasing of the baryon density $\rho_{B}$ due to the presence of hyperons in NS matter decreases the nucleon and lepton fractions according to the requirement of the charge neutrality and $\beta$ equilibium conditions(Eqs.(12) and (13)).
Also, the neutrino emissivities of the reactions B and C are obviously less than that of A, it is due to that they have smaller matrix elements in Eq.(22).
The strongest neutrino emissivity of the reaction A or B is in case (iii), while the weakest one is observed in case (i).
For the reaction C, the neutrino emissivity in case (iii) is less than the corresponding values in cases (i) and (ii) firstly and then increases slowly, equals or exceeds the values in cases (i) and (ii).
While the reactions D-F would never happen within stable NSs, because their threshold densities are larger than the center densities of maximum mass NSs.
As shown in Figs. 2 and 3, the mass ranges of the reactions B and C in case (iii) are $1.671-2.141$M$_{\odot}$ and $1.888-2.141$M$_{\odot}$ for the GM1 model, $1.579-2.038$M$_{\odot}$ and $1.849-2.038$M$_{\odot}$ for TM1 model, respectively.
Next we mainly discuss the effects of the degrees of freedom of hyperons and reactions B, C on the total neutrino emissivity, neutrino luminosity, energy gap of $^{1}S_{0}$ proton SF and NS cooling in case (iii).
Fig. 5 gives the radial distributions of the total neutrino emissivities of the reactions A, B, C for the GM1 model with 1.98, 2.00, 2.10, 2.12M$_{\odot}$ stars in case (iii).
The radial distributions of the total neutrino emissivity for the same mass stars in npe$\mu$ and nphe$\mu$ matter are almost unchanged when r is relatively large (Part I).
However, the reactions B and C happen in succession as the radius decreases (Part II and III), leading to the radial distributions of the total neutrino emissivities in nphe$\mu$ matter are significantly larger than the values in npe$\mu$ matter.
While the radius regions of the increasing total neutrino emissivities shrink continually with the increasing NS mass.
The situation of the TM1 model is like the above in GM1 model.
Fig. 6 shows the total neutrino luminosity as a function of NS mass m in case(iii).
As seen from Fig. 6, whether hyperons are included or not, the neutrino luminosity increases firstly and then decreases with the increasing of the baryon density $\rho_{B}$. Once the mass of NS reaches a value, one luminosity corresponds to two different NSs.
And the total neutrino luminosities of reactions A, B, C within the mass range $1.603-2.067$M$_{\odot}$ and $1.515-1.840$M$_{\odot}$ will be larger than the values in npe$\mu$ matter for the GM1 and TM1 models, respectively.
Fig.7 shows the $^{1}S_{0}$ proton SF critical temperatures as a function of the baryon density $\rho_{B}$ for case(iii) in npe$\mu$ matter(solid lines) and nphe$\mu$ matter(dashed lines), respectively.
One can find that whether or not the NS core appears hyperons, the $^{1}S_{0}$ proton SF critical temperature first increases, reaches a peak and then decreases to zero as the baryon density $\rho_{B}$ increases.
While when the NS core appears hyperons, the $^{1}S_{0}$ proton SF critical temperatures are first below and then above the corresponding values in npe$\mu$ matter within the density ranges of $\rho_{B}=0.0-0.454$ fm$^{-3}$($\rho_{B}=0.0-0.418$ fm$^{-3}$ for the TM1 model) and $\rho_{B}\geq0.454$ fm$^{-3}$($\rho_{B}\geq0.418$ fm$^{-3}$ for the TM1 model) for the GM1 model, respectively.
It is because the total contribution of the Fermi momentum and single-particle energy of protons (see Eqs.(11) and (27)) results in $T_{CP}$ change.
Besides, the density range of the $^{1}S_{0}$ proton SF is widened due to the inclusion of hyperons.
The range of $^{1}S_{0}$ proton SF can achieve coverage or partial coverage in the cores of NSs, which is highly relevant to the reactions A, B.
That is, the presence of hyperons affects not only the reaction A and $^{1}S_{0}$ proton SF critical temperature, but also the NS cooling.
In Figs. 8 and 9, the cooling curves of NSs are calculated through solving the cooling Eq.(24) by assuming isothermal NS cores for case(iii).
Observational data of 8 isolated NSs whose effective surface temperatures have been measured or constrained are listed as compared with the theoretical cooling curves \cite{Slane:APJ616/2004,Zavlin:astro-ph2007,Zavlin:APJ665/2007,Halpern:APJ612/2004,Pavlov:APJ552/2004,McGowan:APJ600/2004,Zavlin:MSAI75/2004,Possenti:AA313/1996,Kargaltsev:APJ625/2005,Ho:APJ375/821}.
As you can see from Figs. 8 and 9, the cooling curves of moderate mass NSs can explain the observational data, while the cooling curves for massive NSs due to the low surface temperature are difficult to explain the existing observational data.
And the cooling curves decrease smoothly with the increasing NS mass, meaning that the cooling of massive NSs undergoes faster neutrino cooling whether the reactions B and C appear in NSs.
In addition, the same mass stars with the reactions B, C(dashed lines) are colder than that in npe$\mu$ matter(solid lines).
Taking case (iii)in GM1 model as an example, we can see that the neutrino luminosity of 1.70, 1.95, 2.03 M$_{\odot}$ NSs in nphe$\mu$ matter are greater than the corresponding values in npe$\mu$ matter which is because the three NSs within the mass range $1.671-2.067$M$_{\odot}$, see Fig. 6 for details.
While the $^{1}S_{0}$ proton SF critical temperatures $T_{CP}$ of 1.70, 1.95, 2.03 M$_{\odot}$ in nphe$\mu$ matter is lower than the value in npe$\mu$ matter(see Figs. 2 and 7 for details), in the case we studied here.
As a result, the reactions A, B are suppressed in advance in nphe$\mu$ matter.
However, the reaction C is not affected by the $^{1}S_{0}$ proton SF.
Therefore, although the neutrino emissivities of the reactions A and B are suppressed with the presence of the $^{1}S_{0}$ proton SF, the total contribution of reactions A, B, C can still speed up a massive NS cooling.
For the TM1 model, the situation is similar to the described in GM1 model.
The cooling curves presented here are estimates in order to make the effects of the $^{1}S_{0}$ proton SF on the reactions A and B be more clearly, in particular, the effects of NS crust and hyperons SF are not considered.
\section{\label{sec:con} Conclusion}
We have studied the effects of the degrees of freedom of hyperons and reactions B, C on the reaction A in NS matter using the two popular RMF parameter sets, GM1 and TM1, respectively.
Firstly,  we used the SU(3) flavor symmetry to obtain the relatively stiff EOS for supporting the observed massive PSR J1614-2230 and J0348+0432.
Secondly, the total neutrino luminosity of the reactions A, B, C are calculated in npe$\mu$ and nphe$\mu$ matter, respectively.
We found that the presence of the reactions B and C caused of the total neutrino luminosity higher than the corresponding values in npe$\mu$ matter within the mass range $1.603-2.067$M$_{\odot}$ for the GM1 model and $1.515-1.840$M$_{\odot}$ for the TM1 model.
Finally, our main purpose is to test the effects of the $^{1}S_{0}$ proton SF on the reactions A, B by comparing cooling curves with observed data.
We will analyze the effects of other baryon SF on the corresponding baryon direct Urca processes in future work.
Our results showed that the cooling rate of the same mass of two NSs with the reactions B and C are obviously faster than that without the reactions B and C.
These features maybe can help to prove the presence of hyperons in massive NSs cores.
%

\begin{acknowledgments}
This work is funded by the National Natural Science Foundation of
China (Grant Nos. 11447165, 11373047, U14311121, 11265009) and Youth Innovation Promotion Association, CAS (Grant Nos. 2016056).
\end{acknowledgments}


\begin{thebibliography}{0}%
\makeatletter
\providecommand \@ifxundefined [1]{%
 \@ifx{#1\undefined}
}%
\providecommand \@ifnum [1]{%
 \ifnum #1\expandafter \@firstoftwo
 \else \expandafter \@secondoftwo
 \fi
}%
\providecommand \@ifx [1]{%
 \ifx #1\expandafter \@firstoftwo
 \else \expandafter \@secondoftwo
 \fi
}%
\providecommand \natexlab [1]{#1}%
\providecommand \enquote  [1]{``#1''}%
\providecommand \bibnamefont  [1]{#1}%
\providecommand \bibfnamefont [1]{#1}%
\providecommand \citenamefont [1]{#1}%
\providecommand \href@noop [0]{\@secondoftwo}%
\providecommand \href [0]{\begingroup \@sanitize@url \@href}%
\providecommand \@href[1]{\@@startlink{#1}\@@href}%
\providecommand \@@href[1]{\endgroup#1\@@endlink}%
\providecommand \@sanitize@url [0]{\catcode `\\12\catcode `\$12\catcode
  `\&12\catcode `\#12\catcode `\^12\catcode `\_12\catcode `\%12\relax}%
\providecommand \@@startlink[1]{}%
\providecommand \@@endlink[0]{}%
\providecommand \url  [0]{\begingroup\@sanitize@url \@url }%
\providecommand \@url [1]{\endgroup\@href {#1}{\urlprefix }}%
\providecommand \urlprefix  [0]{URL }%
\providecommand \Eprint [0]{\href }%
\providecommand \doibase [0]{http://dx.doi.org/}%
\providecommand \selectlanguage [0]{\@gobble}%
\providecommand \bibinfo  [0]{\@secondoftwo}%
\providecommand \bibfield  [0]{\@secondoftwo}%
\providecommand \translation [1]{[#1]}%
\providecommand \BibitemOpen [0]{}%
\providecommand \bibitemStop [0]{}%
\providecommand \bibitemNoStop [0]{.\EOS\space}%
\providecommand \EOS [0]{\spacefactor3000\relax}%
\providecommand \BibitemShut  [1]{\csname bibitem#1\endcsname}%
\let\auto@bib@innerbib\@empty
\end{thebibliography}%


\begin{thebibliography}{}
\bibitem{Yakovlev:APAA42/2004}D. G. Yakovlev, C. J. Pethick,
Ann. Rev. Astron. Astrophys~{\bf42}, 169(2004).

\bibitem{Yakovlev:AIPC983/2008}D. G. Yakovlev et al.,
AIP Conf. Series~{\bf983}, 379(2008).
\bibitem{Ji:PRD57/1998}C. R. Ji, D. P. Min,
Phys. Rev. D~{\bf57}, 5963(1998).

\bibitem{Yakovlev:PU42/1999}D. G. Yakovlev, K. P. Levenfish, Y. A. Shibanov,
Phys. Uspek~{\bf42}, 737(1999).

\bibitem{Zhao:CSB56/2011}E. G. Zhao, F. Wang,
Chin. Sci. Bull~{\bf56}, 3797(2011).

\bibitem{Gao:APSS334/2011}Z. F. Gao et al.
Ap \& SS~{\bf334}, 281(2011).

\bibitem{Sotani:NPA906/2013}H. Sotani, T. Maruyama, T. Tatsumi,
Nucl. Phys. A~{\bf906}, 37(2013).

\bibitem{Schaab:APJ504/1998}C. Schaab, S. Balberg, J. Schaffner-Bielich,
Astrophys. J~{\bf504}, L99(1998).

\bibitem{Wang:PRC81/2010}Y. N. Wang, H. Shen,
Phys. Rev. C~{\bf81}, 025801(2010).

\bibitem{Xu:RAA15/2015}Y. Xu et al.,
Research in Astron. Astrophys~{\bf15}, 725( 2015).

\bibitem{Tsuruta:1964}S. Tsuruta,
Phd. Thesis, Columbia University(1964).

\bibitem{Flowers:APJ205/1976}E. Flowers, M. Ruderman, P. Sutherland,
Astrophys. J~{\bf205}, 541(1976).

\bibitem{Maxwell:APJ231/1979}O. V. Maxwell,
Astrophys. J~{\bf231}, 201(1979).

\bibitem{Gudmundsson:APJ272/1983}E. H. Gudmundsson, C. J. Pethick, R. I. Epstein,
Astrophys. J~{\bf272}, 286(1983).

\bibitem{Page:APJ394/1992}D. Page, J. H. Applegate,
Astrophys. J~{\bf394}, 17(1992).



\bibitem{Kaminker:AA373/2001}A. D. Kaminker, P. Haensel, D. G. Yakovlev,
Astron. Astrophys~{\bf373}, L17(2001).

\bibitem{Yakovlev:NPA752/2005}D. G. Yakovlev et al.,
Nucl. Phys. A~{\bf752}, 90(2005).

\bibitem{Kouvaris:PRD77/2008}C. Kouvaris,
Phys. Rev. D~{\bf77}, 023006(2008).

\bibitem{Blaschke:PRC85/2012}D. Blaschke, H. Grigorian, D. N. Voskresensky, F. Weber,
Phys. Rev. C~{\bf85}, 022802 (2012).

\bibitem{Beznogov:MNRAS447/2015}M. V. Beznogov,  D. G. Yakovlev,
Mon. Not. R. Astron. Soc~{\bf447}, 1598(2015).



\bibitem{Lattimer:PRL66/1991} J. M. Lattimer, C. J. Pethick, M. Prakash, P. Haensel,
Phys. Rev. Lett~{\bf66}, 2701(1991).

\bibitem{Prakash:APJ390/1992}M. Prakash et al.,
Astrophys. J~{\bf390}, 77(1992).

\bibitem{Haensel:AA290/1994}P. Haensel, O. Y. Gnedin,
Astron. Astrophys~{\bf290}, 458(1994).

\bibitem{Levenfish:AL20/1994}K. P. Levenfish, D. G. Yakovlev,
Astron. Lett~{\bf20}, 43(1994).

\bibitem{Gusakov:AA389/2002}M. E. Gusakov,
Astron. Astrophys~{\bf389}, 702(2002).

\bibitem{Xu:CPL28/2011}Y. Xu et al.,
Chin. Phys. Lett~{\bf28}, 079701(2011).

\bibitem{Xu:CTP56/2011}Y. Xu et al.,
Commun. Theor. Phys~{\bf56}, 521(2011).

\bibitem{Xu:CSB59/2014}Y. Xu et al.,
Chin. Sci. Bull~{\bf59}, 273(2014)

\bibitem{Boguta:NPA292/1977}J. Boguta, A. R. Bodmer,
Nucl. Phys. A~{\bf292}, 413(1977).

\bibitem{Boguta:2PLB106/1981}J. Boguta,
Phys. Lett. B~{\bf106}, 250(1981).

\bibitem{Boguta:PLB120/1983}J. Boguta, H. Stocker,
Phys. Lett. B~{\bf120}, 289(1983).

\bibitem{Boguta:PRL59/1987}W. Pannert, P. Ring, J. Boguta,
Phys. Rev. L~{\bf59}, 2420(1987).

\bibitem{Schaffner:PRC53/1996}J. Schaffner, I. N. Mishustin,
Phys. Rev. C~{\bf53}, 1416(1996)

\bibitem{Yang:PRC77/2008}F. Yang, H. Shen,
Phys. Rev. C~{\bf77}, 025801(2008).

\bibitem{Xu:CPL30/2013}Y. Xu et al.,
Chin. Phys. Lett~{\bf30}, 129501(2013).

\bibitem{Demorest:Nature467/2010}P. B. Demorest et al.,
Nature~{\bf467}, 1081(2010).

\bibitem{Antoniadis:Science340/2013}J. Antoniadis et al.,
Science~{\bf340}, 448 (2013).

\bibitem{Weissenborn:PRC86/2012}S. Weissenborn, D. Chatterjee, J. Schaffner-Bielich,
Phys. Rev. C~{\bf85}, 065802(2012).

\bibitem{Miyatsu:PRC88/2013}T. Miyatsu, M. K. Cheoun, K. Saito,
Phys. Rev. C~{\bf88}, 015802(2013).

\bibitem{Weissenborn:NPA914/2013}S. Weissenborn, D. Chatterjee, J. Schaffner-Bielich,
Nucl. Phys. A~{\bf914}, 421(2013).

\bibitem{Lopes:PRC89/2014}L. L. Lopes,  D. P. Menezes,
Phys. Rev. C~{\bf89}, 025805(2014).

\bibitem{Takatsuka:NPA738/2004}T. Takatsuka, R. Tamagaki,
Nucl. Phys. A~{\bf738}, 387(2004).

\bibitem{Xu:CPL29/2012}Y. Xu, et al.,
Chin. Phys. Lett~{\bf29}, 059701(2012).

\bibitem{Oppenheimer:PR55/1939}J. R. Oppenheimer, G. M. Volkoff,
Phys. Rev~{\bf55}, 374{1939}.

\bibitem{Tolman:PR55/1939}R. C. Tolman,
Phys. Rev~{\bf55}, 364(1939).

\bibitem{Leinson:PLB518/2001}L. B.Leinson, A. P{\'e}rez,
Phys. Lett. B~{\bf518}, 15(2001).

\bibitem{Leinson:NPA707/2002}Leinson, L. B.,
Nucl. Phys. A~{\bf707}, 543(2002).

\bibitem{Sprung:NPA168/1971}D. W. L. Sprung, P. K. Banerjee,
Nucl. Phys. A~{\bf168}, 273(1971).

\bibitem{Amundsen:NPA437/1985}L. Amundsen, E. {\O}stgaard,
Nucl. Phys. A~{\bf437}, 487 (1985).

\bibitem{Nishizaki:PTP86/1991}S. Nishizaki, T. Takatsuka, N. Yahagi, J. Hiura,
Prog. Theor. Phys~{\bf86}, 853(1991).

\bibitem{Wambach:NPA555/1993}J. Wambach,  T. L. Ainsworth, D. Pines,
Nucl. Phys. A~{\bf555}, 128(1993).

\bibitem{Slane:APJ616/2004}P.Slane et al.,
Astrophys. J~{\bf616}, 403(2004).

\bibitem{Zavlin:astro-ph2007}V. E. Zavlin,
astro-ph/0702426(2007).

\bibitem{Zavlin:APJ665/2007}V. E.Zavlin,
Astrophys. J~{\bf665}, L143(2007).

\bibitem{Halpern:APJ612/2004}J. P. Halpern et al.,
Astrophys. J~{\bf612}, 398(2004).

\bibitem{Pavlov:APJ552/2004}G. G. Pavlov et al.,
Astrophys. J~{\bf552}, 129(2001).

\bibitem{McGowan:APJ600/2004}K. E. McGowan et al.,
Astrophys. J~{\bf600}, 343(2004).

\bibitem{Zavlin:MSAI75/2004}V. E. Zavlin, G. G. Pavlov,
Mem. Soc. Astron. Ital~{\bf75}, 458(2004).

\bibitem{Possenti:AA313/1996}A. Possenti, S. Mereghetti, M. Colpi,
Astron. Astrophys~{\bf313}, 565(1996).

\bibitem{Kargaltsev:APJ625/2005}O. Y. Kargaltsev et al.,
Astrophys. J~{\bf625}, 307(2005).

\bibitem{Ho:APJ375/821}W. C. G. Ho et al.,
Astrophys. J~{\bf375}, 821(2007).

\end{thebibliography}
\end{document}